\documentclass[12pt]{article}

\usepackage{times}
\usepackage{xcolor}
\usepackage{graphicx}
\usepackage{lineno}

\newenvironment{sciabstract}{%
\begin{quote} \bf}
{\end{quote}}

\title{Network Structure shapes the Impact of Diversity in Collective Learning}

\author
{Fabian Baumann, Agnieszka Czaplicka, and Iyad Rahwan$^*$\\
\\
\normalsize{Center for Humans and Machines,}\\
\normalsize{Max Planck Institute for Human Development, Lentzeallee 94, Berlin 14195, Germany.}\\
\normalsize{Corresponding author: rahwan@mpib-berlin.mpg.de}
}

\date{}

\begin{document} 

\baselineskip24pt

\maketitle 
\begin{sciabstract}
It is widely believed that diversity arising from different skills enhances the performance of teams, and in particular, their ability to learn and innovate. 
However, diversity has also been associated with negative effects on the communication and coordination within collectives. Yet, despite the importance of diversity as a concept, we still lack a mechanistic understanding of how its impact is shaped by the underlying social network.
To fill this gap, we model skill diversity within a simple model of collective learning and show that its effect on collective performance differs depending on the complexity of the task and the network density.
In particular, we find that diversity consistently impairs performance in simple tasks. In contrast, in complex tasks, link density modifies the effect of diversity: 
while homogeneous populations outperform diverse ones in sparse networks, the opposite is true in dense networks, where diversity boosts collective performance.
Our findings also provide insight on how to forge teams in an increasingly interconnected world: the more we are connected, the more we can benefit from diversity to solve complex problems.
\end{sciabstract}

\section*{Introduction}
Living in an increasingly interconnected world humans communicate with higher rates than ever before \cite{gil2011mediating}. 
With the goal to optimize collective problem-solving, previous studies have focused on the effects of network structure on the overall performance of a population in different tasks.  
Counter intuitively, it was found that groups do not necessarily benefit from densely connected networks, where information can be exchanged freely and spreads rapidly \cite{lazer2007network,barkoczi2016social,mason2012collaborative,derex2016partial,derex2018divide,mason2008propagation,derex2020cumulative,migliano2020hunter,cantor2021social}.
In addition to enhanced communication, increasing global connectivity also means that individuals from different backgrounds and cultures interact and work together to solve problems collectively  \cite{saxena2014workforce}. 
{It is therefore also crucial to investigate how individual differences within networked groups or societies -- often referred to as diversity -- shape the success of collective social endeavors.}

Empirical studies have found contradictory, i.e., positive \textit{and} negative, effects of diversity on the performance of groups and larger organisations \cite{bell2011getting,alesina2002trusts,schimmelpfennig2022paradox,pescetelli2021modularity,horwitz2007effects,weber2003cultural,steiner1972group,converse1993shared,cronin2007representational,putnam2007pluribus,eberle2020ethnolinguistic,coles2020director,reagans2001networks,centola2022network}.
For instance, an increased professional diversity predicts greater productivity \cite{moro2021universal,alesina2002trusts} of cities. 
Furthermore, within organizations, the teams with the highest diversity in educational backgrounds have been found to be the ones that are most innovative \cite{bell2011getting}.
It was also shown that diverse teams are less prone to detrimental groupthink \cite{coles2020director}. 
However, there is also evidence that diversity can negatively impact collective performance through various adverse effects. 
Specifically, diversity may impair the ability of groups to communicate and coordinate \cite{weber2003cultural,steiner1972group,converse1993shared,cronin2007representational}, be associated with reduced trust between individuals \cite{putnam2007pluribus,alesina2002trusts}, and can be a source for social tension and conflict \cite{schimmelpfennig2022paradox,eberle2020ethnolinguistic}. 
{From a theoretical perspective, there is limited understanding of how network structure and diversity affect collective performance, as previous modelling efforts have focused on homogeneous populations of identical individuals \cite{lazer2007network,watts1998collective,barkoczi2016social}.}

\begin{figure}\begin{center}\includegraphics[width=\linewidth]{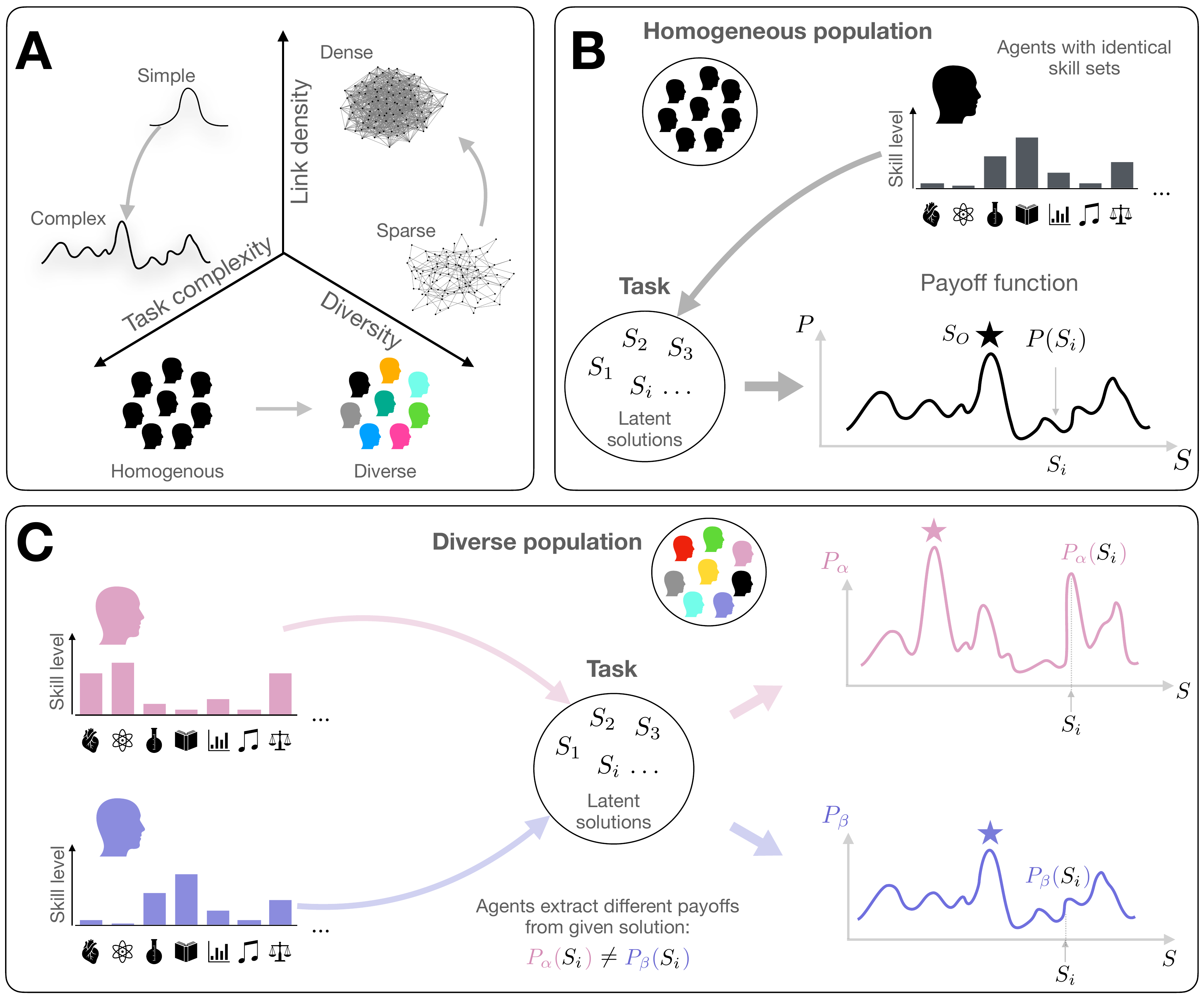}
\caption{\textit{Schematic overview of the model.} Panel A depicts the features of the model that are varied including (i) task complexity, (ii) link density of the underlying network, and (iii) the composition of the population in terms of skill diversity.
In panels B and C, we schematically depict how agents' skills map onto payoff functions $P$.
In a homogeneous population, all agents have the same set of skills (panel B). 
This results in a \textit{single} payoff function, which maps a latent solution $S_i$ to payoff value $P(S_i)$ an agent is able to extract from it. 
In homogeneous populations, all agents can extract the same payoff from a given solution $S_i$, and hence there is one optimal solution $S_O$ (star symbol) for all agents.
In panel C, we show the case of a diverse population, where agents generally have different distributions of skills. For brevity, we depict two agents with different skill sets and their corresponding distinct payoff functions $P_\alpha$ and $P_\beta$. 
As depicted two agents in diverse populations will generally not extract the same amount of payoff from a particular solution $S_i$. 
Accordingly, also their optimal solutions differ (shown as star symbols).}\label{fig:schematic} \end{center} \end{figure}

Under which circumstances do networked collectives benefit from diversity? 
To answer this question, we build on established modeling efforts, however, overcome their limitations by considering heterogeneous individuals. In particular, we study {skill} diversity within the paradigm of social learning, which has also been used in empirical studies on collective problem solving including experiments on the evolution of culture \cite{boyd1988culture,derex2016partial,derex2018divide}, and more narrow cases where groups solved specific real world tasks \cite{almaatouq2021task,watts1998collective}. 

From a theoretical perspective, the dynamics of social learning can be mapped onto a collective search process in an underlying solution space \cite{lazer2007network,barkoczi2016social}, where the quality of a solution $S$ is assessed using a payoff function $P(S)$: {the higher the payoff $P(S)$, the better the solution $S$. Importantly, we assume that the payoff function $P$ does not solely result from the collective task itself, but is shaped by the skills of agents. Put differently, an agent's skill set determines how much value $P$ they can extract from $S$, i.e., $P(S)$ corresponds to their contribution to the collective performance of the group. The mapping of task and skills onto a payoff function is schematically depicted in Fig.~\ref{fig:schematic}.}

{To navigate the space of latent solutions efficiently, i.e., to rapidly discover beneficial solutions, models of social learning typically assume that agents balance exploitative and explorative behaviour \cite{laland2004social,lazer2007network,barkoczi2016social,mason2012collaborative,rendell2010copy,fang2010balancing}.}
While exploitation corresponds to copying solutions from other agents, exploration refers to individual learning, where agents discover new solutions by themselves.
The balance between these two types of learning, i.e., individual learning and copying, typically leads to an iterative improvement of both individual and collective performance \cite{laland2004social,barkoczi2016social}. 

Following recent approaches to the study of social learning \cite{derex2016partial,mason2012collaborative,barkoczi2016social,smolla2019cultural,almaatouq2020adaptive,lazer2007network}, we assume that individuals are embedded in a social network and can therefore only copy solutions from a subset of other agents, i.e., their neighbors in the network. 
To copy solutions, agents consider the neighbor which holds the currently best solution and adopt it only if it improves their payoff \cite{barkoczi2016social,mason2012collaborative,lazer2007network}. 
As in previous models, agents innovate locally, i.e., they can only explore solutions that are close to their current ones.
{We quantify the collective performance of a population in terms of the average payoff of agents $\langle P\rangle = N^{-1}\sum_{i=1}^{N} P_i$, where $P_{i}$ corresponds to the contribution of agent $i$. 
Hence, each agent contributes to $\langle P\rangle$ with equal weight. 
In general, $\langle P\rangle$ can take values between 0 and 1, and the higher the value of $\langle P\rangle$, the better is the collective performance.}
For a more detailed description of the model, see Methods. 

In the following, we investigate the collective dynamics along three dimensions by varying (i) task complexity, (ii) network structure, and (iii) group composition with regard to {skill} diversity. The setup is schematically depicted in Fig.~\ref{fig:schematic}A. 

Real-world tasks come in different complexities and it has been shown theoretically as well as empirically that task complexity strongly impacts collective performance \cite{lazer2007network,barkoczi2016social,almaatouq2021task}. 
{The complexity of a task is usually defined by the statistical properties of its payoff function, in particular, the correlations between different solutions \cite{kauffman1987towards,lazer2007network,barkoczi2016social,almaatouq2021task,mason2012collaborative}.}
In this picture, a simple problem is defined by a smooth payoff function with a single peak that corresponds to the optimal solution $S_O$, i.e., the solution with the highest payoff \cite{barkoczi2016social,kauffman1987towards}. 
Social learning is generally well adapted to solve simple problems and groups are usually much faster in finding $S_O$ as compared to isolated problem solvers \cite{barkoczi2016social,lazer2007network}.

Complex tasks are characterized by a rugged payoff landscape. 
While rugged landscapes have an optimal solution $S_O$, complex tasks give rise to additional locally optimal but globally suboptimal solutions to which {collectives} tend to converge, and from where they cannot escape \cite{lazer2007network,barkoczi2016social}. 
This feature complicates the navigation of rugged solution spaces and makes complex tasks harder to solve, both individually and as a group \cite{lazer2007network,almaatouq2021task,barkoczi2016social}.
One of our goals is to investigate if, and how, the complexity of a task modifies the effects of {skill} diversity on the {collective} performance.
{In panel A of Fig.~\ref{fig:schematic}, task complexity is depicted schematically in terms of its associated payoff landscape: a smooth landscape with a single optimum is associated with a simple task and a rugged landscape with multiple (suboptimal) peaks corresponds to a complex task. Following previous works, we use the NK model \cite{kauffman1987towards,lazer2007network,barkoczi2016social} to generate payoff functions for both simple and complex tasks.} For details on the NK model, see Method section.

The effects of network structure on the collective dynamics of homogeneous problem solvers have gained a lot of attention. 
Empirical studies \cite{derex2016partial,watts1998collective} and theoretical models \cite{barkoczi2016social,lazer2007network} have shown that the structure of the communication network strongly affects collective problem solving. 
{An important finding is that it is often not the most densely connected populations that perform best, but rather the partially connected ones, especially when the task to be solved is complex \cite{lazer2007network, derex2016partial, derex2018divide}.}
{Given the crucial role of networks, we study how skill diversity and network structure interact to influence the performance of a group.}
Specifically, we focus on the interplay between link density and diversity and consider random networks with different average degrees that {obey a Poisson degree distribution.}
By that we cover the range from very sparse to more dense networks, including the limiting case of fully connected networks. 

{Overall, our focus is on the dynamics of collective problem solving induced by populations with (increasing) skill diversity in contrast to homogeneous ones.}
As in previous studies, we assume that homogeneous populations are composed of identical agents \cite{lazer2007network,barkoczi2016social,watts1998collective}. 
{By definition, in homogeneous populations all agents possess the same distribution of skills, which results in a single payoff function $P$. 
Each agent therefore extracts the same amount of  value, $P(S)$, from a particular solution $S$, and there is consequently a unique solution $S_O$, which defines the optimum for all agents with $P(S_O)=1$.
The mapping from the task and agents' skills onto the payoff function $P$ is schematically depicted in Fig.~\ref{fig:schematic}B. The optimal solution with $P=1$ is marked by the star symbol.} 

In diverse populations the situation is different. 
{We formalize diversity by assuming that agents generally differ with respect to their skills. 
As depicted for two agents with different skill distributions in panel C of Fig.~\ref{fig:schematic}, this translates into distinct payoff functions ($P_\alpha$ and $P_\beta$) that the agents use to navigate the space of latent solutions.
It generally holds that $P_\alpha(S_i)\neq P_\beta(S_i)$ for a particular solution $S_i$, i.e., the two agents extract different payoff values from a particular solution.}
It is a direct consequence of diverse skill sets that there is no well-defined optimal solution $S_O$. 
Instead, different {skill sets} map to different solutions, from which agents extract the highest payoff (star symbols). 
{In multi-disciplinary teams agents should focus on solutions that match their abilities. For instance, an individual with high analytical abilities is expected to extract more payoff from a solution associated with a mathematical issue, while a co-worker who is more strong in the field of linguistics should focus on parts of the task that are connected to language in order to maximize their contribution to the collective performance.}

{If not denoted otherwise,} we consider populations of $N=1000$ agents. 
We assume that diverse populations can be subdivided into equally sized classes of agents with identical {skill distributions}.
Thus by increasing the number of skill classes, we increase the level of diversity in a population. Note that the collective performance of homogeneous and diverse populations are indeed comparable as each payoff function is normalized. Hence, we generally find that the optimal collective performance is $\langle P\rangle =1$, independent of the level of diversity.

Consistent with empirical studies \cite{bell2011getting,alesina2002trusts,schimmelpfennig2022paradox,pescetelli2021modularity,horwitz2007effects}, we find strong effects of diversity on {collective} performance, and we demonstrate that those effects are mixed, i.e., groups may benefit but also suffer from {skill} diversity. 
In the model, positive and negative effects of diversity arise as a result of varying the task complexity and the network structure. 
Diversity generally reduces the performance of groups in simple tasks. 
In the case of complex tasks, the link density of the network modifies the effect of diversity: the more we are connected, the more we can benefit from diversity to solve complex problems.

\begin{figure}\begin{center}\includegraphics[width=\linewidth]{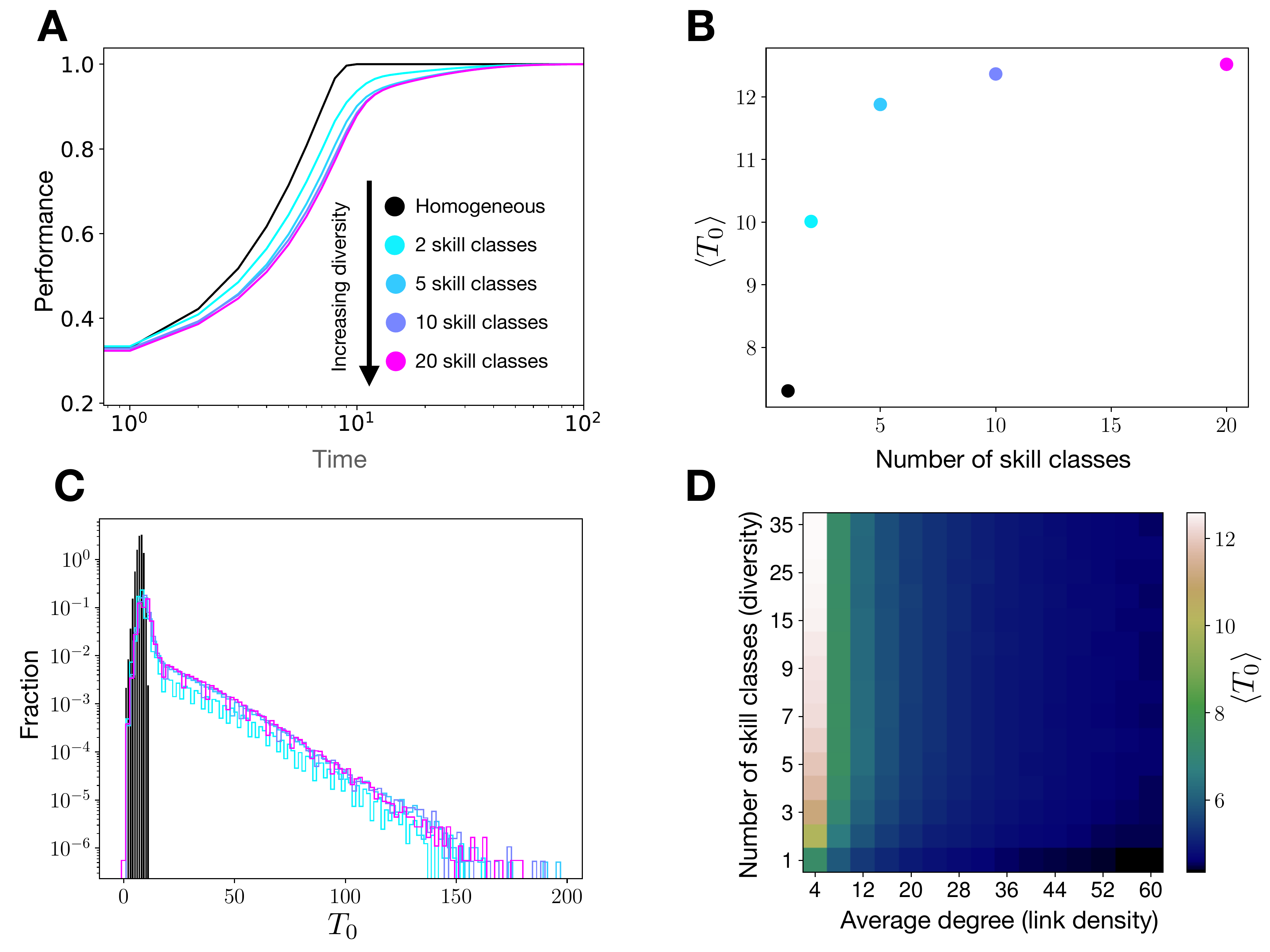}
\caption{\textit{Diversity in simple tasks}. 
Panels A-C show the average payoffs over time (A), average times $\langle T_O\rangle$ to discover the globally optimal solutions (B), and the normalized histograms of $T_O$ (C) for homogeneous and (increasingly) diverse populations on sparse networks with an average degree of $\langle k\rangle=4$. 
Panel D depicts $\langle T_O\rangle$ (color coded) as a function of the average degree of the underlying network and the level of diversity. The reported results are averages over 2500 realizations.} \label{fig:simpleLandscape}\end{center}\end{figure}

\section*{Results}
\subsection*{Diversity in simple tasks}
Figure~\ref{fig:simpleLandscape} shows results obtained for {simple tasks}. In panels A-C, we consider populations coupled via sparse random networks with an average degree of $\langle k\rangle=4$, and colors encode different levels of diversity. In panel D, we vary both the level of diversity and the average degree of the underlying network.

{Figure~\ref{fig:simpleLandscape}A depicts the collective performance quantified as the average payoff $\langle P\rangle$ over time.}
{We find that all populations, both homogeneous (black) and diverse (color coded) ones, eventually converge to the solutions $S_O$ with maximum payoff,  $P(S_O)=1$.
However, the time to reach $S_O$, i.e.,  to extract the maximum payoff collectively depends on the level of diversity.}

To quantify the observed effect of diversity further, we consider the times, $T_O$, that agents need to converge to optimal solutions. 
Figure~\ref{fig:simpleLandscape}B depicts the {mean time} $\langle T_O\rangle$ after which individual agents encounter $S_O$. {Clearly, $\langle T_O\rangle$ is smallest in homogeneous populations}, increases with diversity, and saturates for high levels of diversity.

In Fig.~\ref{fig:simpleLandscape}C, we show the distributions of $T_O$ for the different levels of diversity. 
While in the case of homogeneous populations the distribution is narrowly centered around a small value, the distributions are much broader and characterized by a pronounced positive skew for diverse populations. 
Note, that the qualitative change in shape of the distribution already emerges for low levels of diversity, i.e., in the case of two different skill classes (cyan line).

Finally, Fig.~\ref{fig:simpleLandscape}D shows $\langle T_O\rangle$ as a function of diversity and the link density of the network, which is tuned by the value of the average degree $\langle k\rangle$. 
While for increasing $\langle k\rangle$, the effect of diversity becomes less pronounced the qualitative behavior is the same: as populations become more diverse, the average time $\langle T_O\rangle$ to reach optimal solutions increases.
{We also find that to solve simple problems collectively} dense networks are generally beneficial as agents converge faster to optimal solutions in networks with larger values of $\langle k \rangle$. Note that this trend can be observed for all levels of diversity, varying along the vertical axis in panel D of Fig.~\ref{fig:simpleLandscape}.

{Taken together, the results reported in Fig.~\ref{fig:simpleLandscape} suggest that in simple tasks diversity hampers collective performance by reducing the speed of collective problem solving. We find that homogeneous populations are most efficient in optimizing collective payoffs, and diverse populations need more time. This (qualitative) conclusion holds independent of the average degree $\langle k\rangle$.} {To complement the findings depicted in Fig.~\ref{fig:simpleLandscape}, in the SM we show results for different types of networks, which show the same qualitative behavior.}

\subsection*{Diversity in complex tasks}
\begin{figure}\begin{center}\includegraphics[width=\linewidth]{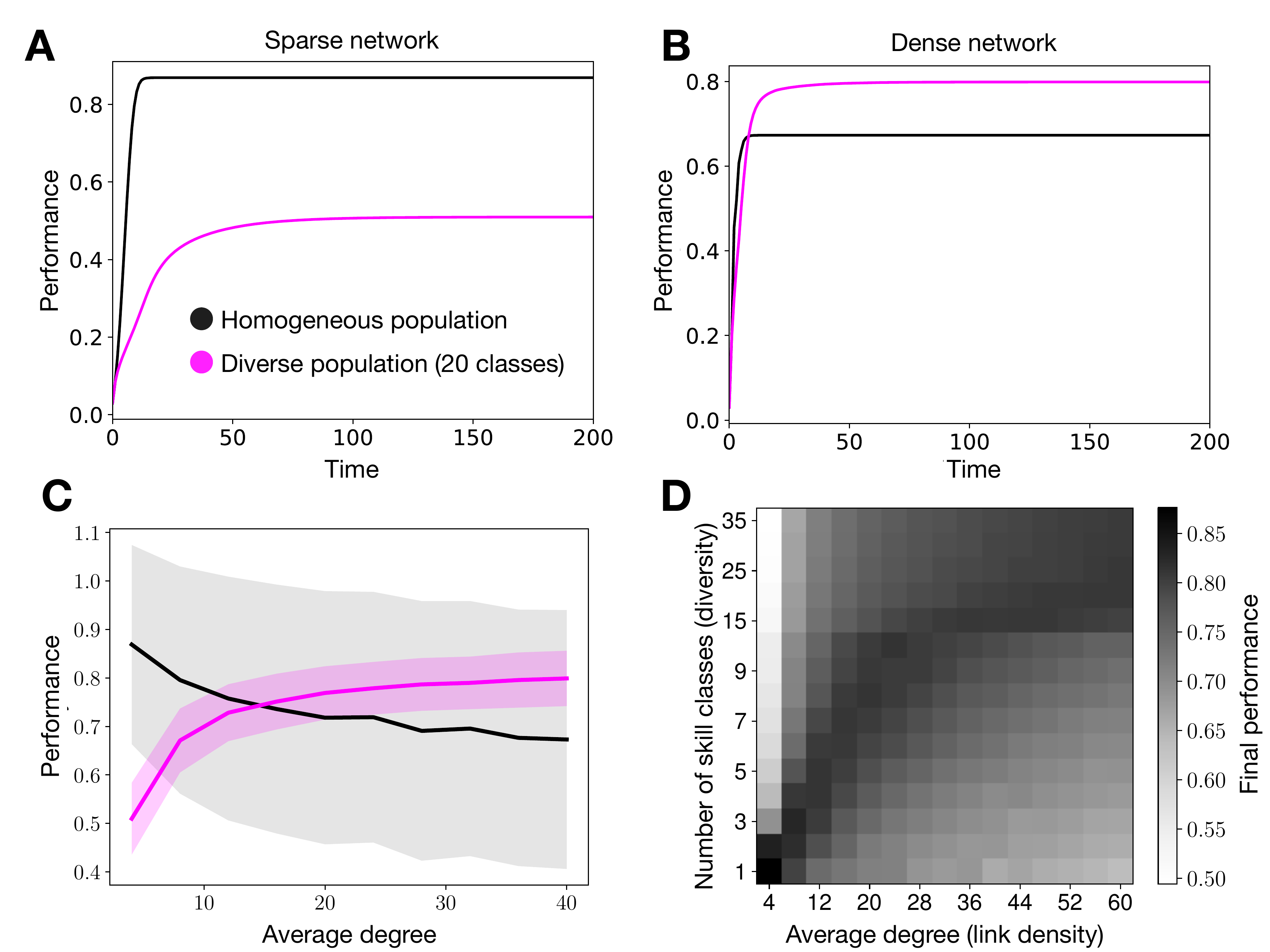}\caption{\textit{Diversity in complex tasks}. 
In panels A and B, we depict average payoffs over time of homogeneous and diverse populations on sparse ($\langle k\rangle=4$) and more dense ($\langle k\rangle=40$) random networks, respectively.
Panel C depicts the average final payoffs for increasing average degrees of the underlying network. 
Panel D shows the average final payoffs (in gray scale) as a function of the average degree of the underlying network and the level of diversity.
The reported results are averages over 2500 realizations. The shaded areas in panel C depict standard deviations.}\label{fig:complexProblem}\end{center}\end{figure}

Figure~\ref{fig:complexProblem} summarizes the results for complex tasks. 
In particular, in panels A-C, we compare {the collective dynamics and final performance} of homogeneous and diverse populations on sparse and dense {random} networks. In panel D, we systematically vary both the diversity and the average degree of the underlying networks over wider ranges.

In Fig.~\ref{fig:complexProblem}A, we depict the {collective} performance over time on sparse networks with the same average degree as in panels A-C of Fig.~\ref{fig:simpleLandscape}, i.e., of $\langle k\rangle = 4$.
{First, we note that both the homogeneous and diverse populations (20 payoff classes) are not able reach the optimal collective performance $\langle P\rangle=1$.}
Instead, due to the ruggedness of the complex solution space populations {tend to} converge to globally suboptimal peaks, which reduces the performance to $\langle P\rangle<1$.
As in the case of simple tasks, homogeneous and diverse populations converge quickly and more slowly to their {performance level}, respectively.
More importantly, however, we find that the final performance of diverse populations is worse than the one reached by homogeneous populations.
As we will see in the following, this is not always the case and homogeneous populations do not generally outperform diverse ones in complex tasks.
Instead, the effect of diversity on collective performance depends on the link density of the underlying network.

In Fig.~\ref{fig:complexProblem}B, we show results for a random network with the tenfold average degree as compared to panel A, i.e.,  $\langle k\rangle = 40$.
By fixing all other model parameters, we find that the {increased link density} significantly impacts the collective dynamics and qualitatively reverses the previous result.
Although, again, diverse populations converge more slowly, the final collective performance of diverse populations is superior to the one reached by homogeneous populations.
These findings suggest that {collective problem solving in networked populations} exhibits a transition point with respect to link density above which diversity becomes beneficial and boosts performance. 

To visualize this phenomenon more clearly, Fig.~\ref{fig:complexProblem}C depicts the final performance of homogeneous and diverse populations in networks of increasing average degree.
For low average degrees, homogeneous populations outperform diverse ones, i.e., final solutions found by homogeneous populations have on average higher payoffs. 
However, as $\langle k\rangle$ increases, diversity becomes beneficial and the final collective performance reached by diverse populations exceeds the one of homogeneous populations. 
{As we show in the SM, this result holds in the limiting case of a fully connected network, where the final collective performance is lowest for homogeneous populations and improves for higher levels of diversity. } 

In Fig.~\ref{fig:complexProblem}D, we shed further light on the combined effect of link density {and diversity by varying both simultaneously.}
In gray scale we show the final collective performance as a function of the average degree and the number of {skill} classes. 
As we have already learnt from Fig.~\ref{fig:complexProblem}C, the performance of homogeneous populations is impaired as the average degree of the network increases.
Similarly, on sparse networks the introduction of diversity, i.e., the increase of the number of {skill} classes, leads to a strong decrease in performance.
This is not the case for networks with a higher link density (increasing values of $\langle k \rangle$).
Instead, we find that the more dense the network becomes, the higher is the \textit{ideal} level of diversity, {which optimizes the final collective performance of the population.} 
{To demonstrate that these findings are not restricted to random networks with Poisson degree distribution, we performed additional simulation on random networks with fixed degree and small world networks, see SM.}

\section*{Discussion}
While various experimental works have found strong effects of diversity on {collective} performance, theoretical studies that provide a mechanistic understanding of the underlying dynamics are scarce. 
Here, we fill this gap by means of a simple model of social learning that we extend by agents that differ {with respect to their skills.}
In line with previous studies, we find that diversity can be both beneficial and detrimental to {collective} performance \cite{bell2011getting,alesina2002trusts,schimmelpfennig2022paradox,pescetelli2021modularity,horwitz2007effects,weber2003cultural,steiner1972group,converse1993shared,cronin2007representational,putnam2007pluribus,eberle2020ethnolinguistic,coles2020director,hong2004groups}.  
Our modeling approach allows to draw intuitive conclusions on the effects of diversity that are related to the underlying network structure and the complexity of the task. 

{For simple problems, all populations -- regardless of their level of diversity -- finally reach the optimal collective performance.
To gauge the effect of diversity, we have quantified how much time it takes for populations to converge to $\langle P\rangle =1$. 
Diversity generally slows down the speed of convergence optimal performance and therefore hampers collective problem solving in terms of efficiency.}
While this effect is more pronounced on sparse networks, it holds qualitatively true on densely connected networks.

In the case of complex tasks, the situation is more differentiated. There is no constant effect of diversity on collective problem solving.
{Instead, the effects of skill diversity depend on the underlying network structure.} 
{On sparse networks with a low average degree, diversity impairs collective problem solving and diverse populations perform worse as compared to homogeneous ones.} 
However, as soon as networks become sufficiently dense diversity becomes beneficial and can boost collective performance beyond that of homogeneous populations. 

There are two separate effects {introduced by skill diversity} that allow to qualitatively understand these findings.
{First, in diverse populations a particular solution $S$ is not evaluated equally by all agents. Thus, a solution may be profitable for some agents (they can extract a high payoff value from it), but not for others (low payoff): we generally find $P_\alpha(S)\neq P_\beta(S)$ as depicted in panel B of Fig.~\ref{fig:schematic}.
This feature of diversity introduces noise to the dynamics of collective problem solving. It negatively affects the collective filtering and efficient dissemination of solutions throughout the population.}

In the case of complex tasks, there is a second effect of diversity that results from the ruggedness of the payoff landscapes.
{It is not only that two agents with different skill distributions extract different payoff values from a particular solution $S$. 
Additionally, the local neighborhood of a solution in the payoff space may differ.
Indeed, while $S$ may correspond to a local peak, or maximum, in the payoff function of one agent where they can get stuck, $S$ potentially corresponds to a local minimum for a differently skilled agent. The second agent is then able to improve upon the solution by local exploration.} 
This feature gives rise to a mechanism that helps agents in diverse populations to escape suboptimal solutions to complex task. 
In other words, precisely because two agents have different {skill sets}, they can mutually benefit from sharing information to improve their solutions.

In simple tasks, the payoff landscape is smooth and agents are able to reach solutions that lead to optimal performance, even by individual and local exploration only \cite{levinthal1997adaptation}. 
{Copying from peers with identical skills accelerates this process} as time resources spent on individual learning are saved \cite{laland2004social}. 
In essence, solutions are evaluated equally by all agents and homogeneous populations can efficiently filter and disseminate beneficial solutions.
{The same process is disrupted in diverse populations, as agents with different skills do not agree on the value of a particular solution ($P_\alpha(S)\neq P_\beta(S)$), ultimately decreasing the convergence rate towards optimal collective performance $\langle P \rangle=1$.}
The finding is in line with empirical studies, suggesting that diversity often reduces the ability of groups to communicate efficiently, thereby creating an obstacle for collective problem solving in simple tasks \cite{schimmelpfennig2022paradox}. 

In complex tasks, the situation is different and diversity can boost collective performance:
diverse populations outperform homogeneous ones on densely connected networks.
As argued previously, homogeneous populations rapidly collapse to, and get trapped at, suboptimal solutions when they copy solutions from their best performing peers \cite{lazer2007network,barkoczi2016social}.
{Noise from various sources, originating either from different social learning strategies, less well-connected networks or distrust between individuals, has been shown to lead to better exploration of the solution space and improved performance \cite{barkoczi2016social,mason2012collaborative,lazer2007network,massari2019distrust}.}
Here we identify diversity as another way to boost collective performance. 
{Different skill sets that map onto distinct payoff functions allow agents to mutually assist each other in escaping local optima, a benefit of diversity that is based on the information exchange among dissimilar peers.}

While our work provides a novel perspective on the relationship between diversity, network structure, and collective performance it comes with noteworthy limitations.
We considered the specific type of diversity, where the population can be subdivided in equally sized {skill} classes.
Generally, however, it can be assumed that groups of any kind are not of the same size.
As previously shown for different types of social dynamics ranging from collective opinion change to the emergence of conventions a broader distribution of group sizes can affect system level outcomes \cite{centola2018experimental,civilini2021evolutionary}. 
An interesting extension to our presented results could consider different task complexities simultaneously. More specifically, our framework allows to investigate cases where a particular task is complex for most agents in the population, while others can solve it easier. This may shed light on the role of diversity in the context of generalists and specialists, which has previously been studied for homogeneous populations \cite{smolla2019cultural}.

Overall, our findings suggest that diversity offers a pathway to boost collective performance in complex tasks. 
Specifically, we have shown that the more we are connected, the more we can benefit from diversity to solve complex problems -- a finding which informs the compilation of problem solving teams in an increasingly interconnected and diverse world.

\section*{Methods}
\subsection*{Dynamics of social learning}
We consider a populations of $N$ agents, which are embedded in undirected networks.
In each discrete time step $t$, agent $i$ aims to improve their solution in a two-step procedure. 
First, they attempt to learn, or copy, a solution from a neighbour in the social network.
As in previous studies (e.g. Refs. \cite{lazer2007network,barkoczi2016social,mason2012collaborative}), we implement a copying strategy, where agents follow their best neighbor, defined as the agent that currently holds the best solution, i.e., the solution that gives rise to the highest payoff for agent $i$.
If the solution improves the current payoff of agent $i$, they adopt it. 
Otherwise, agent $i$ switches to innovation, i.e. they locally explore the solution landscape in search of a new, unknown solution. To innovate a new solution by local exploration agents modify a single digit of their current solution.
As in the case of copying, the innovated solution is only adopted if it increases the agent's current payoff \cite{csaszar2010much,lazer2007network,kauffman1987towards,barkoczi2016social}.

To evaluate solutions, and to compare them to each other, agents use a payoff function $P$, which maps a solution $S$ to a numerical value $P(S)$ that lies between zero and one: the higher $P$, the more value an agent can extract, i.e., the better the solution.
In homogeneous populations all agents utilize the same payoff function $P$ and therefore evaluate each solution equally. 
Instead, in a diverse population agents use different payoff functions. 
A particular solution is evaluated differently by two agents $i$ and $j$, which belong to payoff classes $\alpha$ and $\beta$, respectively, as we generally find $P_\alpha(S)\neq P_\beta(S)$.

\subsection*{Generating payoff functions}
To generate payoff landscapes corresponding to varying task complexities we use the NK model \cite{kauffman1987towards}. 
In its original version the model is fully defined by two parameters, $N_\mathrm{NK}$ and $K_\mathrm{NK}$. 
Each solution $S$ in the problem space can be regarded as a vector of $N_\mathrm{NK}$ bits, or elements, and has a well defined payoff value $P$. The total number of possible solutions is thus given by $2^{N_\mathrm{NK}}$. The payoff contribution of a particular solution $S$ is defined as the mean payoff contribution of each element, which is a random number drawn from a uniform distribution on the interval between 0 and 1. For $K_\mathrm{NK}=0$, the average is simply taken over the contribution of each element $b_i$, i.e. we have $P=N^{-1}\sum_i b_i$. For $K_\mathrm{NK}>0$, the contribution of each element $f(b_i\vert b_i, b_{i+1}, \dots,b_{K_\mathrm{NK}})$ is determined by $K_\mathrm{NK}$ additional and independent elements. The total payoff is then computed as $P=N^{-1}\sum_i f(b_i\vert b_i, b_{i+1}, \dots,b_{K_\mathrm{NK}})$, where the $K_\mathrm{NK}$ elements are determined randomly.

Following previous studies, we first normalize the payoff function $P$ by its maximum value $P_\mathrm{max}$ \cite{barkoczi2016social,lazer2007network,siggelkow2005speed}.
For intermediate values of $K_\mathrm{NK}$ the distribution of payoffs tends towards a normal distribution, such that most solutions have similar payoffs, which makes it hard to distinguish collective performance in different conditions \cite{barkoczi2016social}. 
We therefore take the normalized payoff function to the power of 8, such that there are only very few good solutions, as was done in \cite{barkoczi2016social,lazer2007network}. We distinguish between two scenarios: simple and complex tasks for which we set the parameters of the NK model to $(N_\mathrm{NK}, K_\mathrm{NK}) = (15,0)$ and $(N_\mathrm{NK}, K_\mathrm{NK})= (15, 7)$, respectively.

\section*{Code availability}
Simulation code is available at https://github.com/naibaf2/diversity

\bibliography{scibib}
\bibliographystyle{unsrt}

\newpage
\section*{Supplemental Information}
In this Supplemental Information, we report results which demonstrate that our main findings on the effects of skill diversity hold for varied system sizes and different types of networks.

\subsection*{Simple tasks}
\begin{figure}
\includegraphics[width=\linewidth]{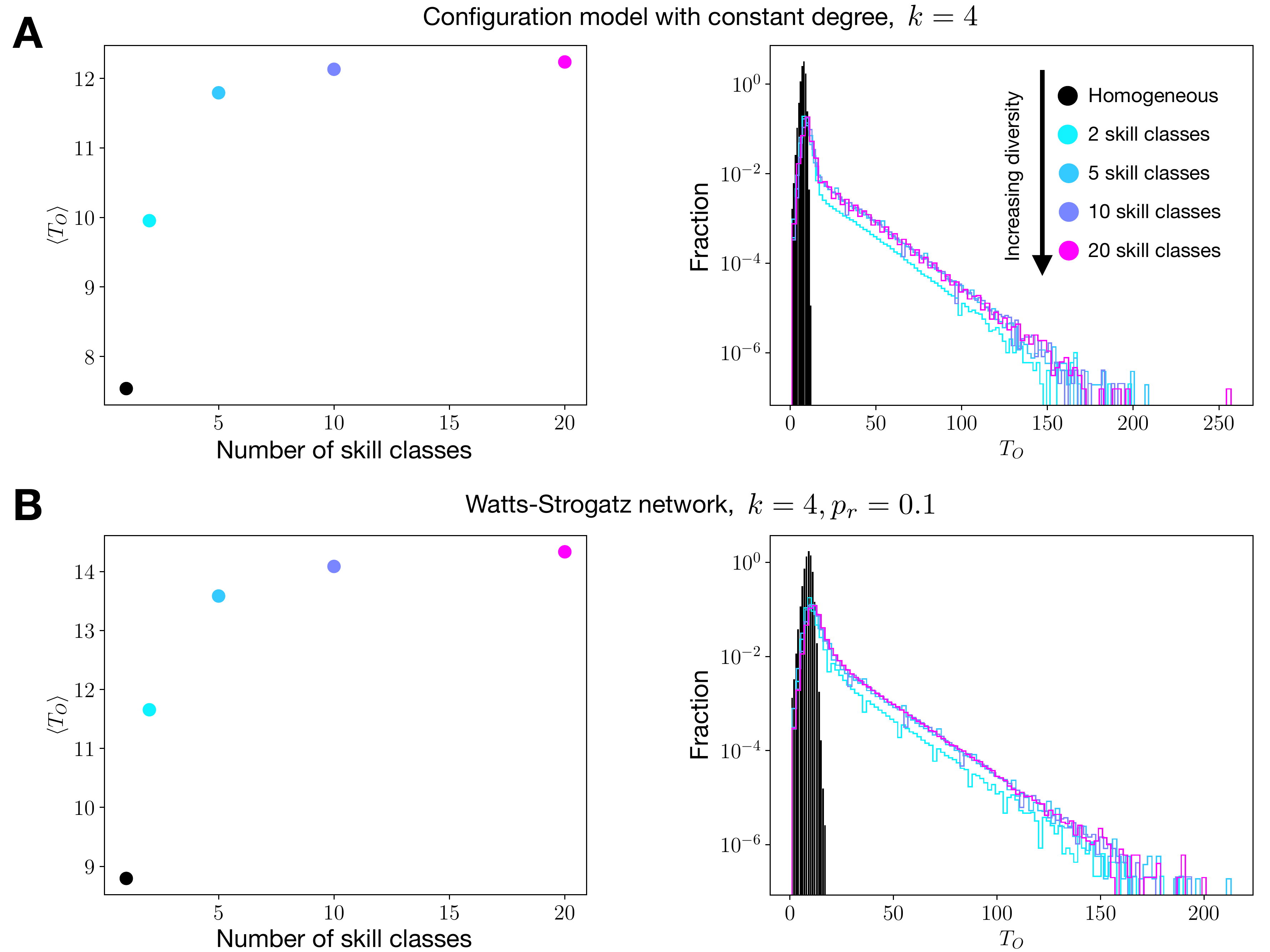}
\caption{Collective performance in simple tasks. Mean times $\langle T_0\rangle$ to reach optimal performance (left) and the corresponding distributions of $T_0$ for different levels of diversity (right). 
In panel A, we show results for random networks generated by the configuration model with a fixed degree of $k=4$, i.e. each node has four adjacent nodes. 
In panel B, we show results for small world networks generated by the Watts-Strogatz model, where each link of a regular grid network is rewired with probability $p_r$. 
Here we chose a regular gird network where each node is connected to its $k=4$ nearest neighbors and set the rewiring probability to $p_r=0.1$. In all shown cases the system size is $N=1000$. 
The results are averaged over 2500 realizations.} \label{fig:figure_simpleLandscape_SM}
\end{figure}

In Fig.~\ref{fig:figure_simpleLandscape_SM}, we show results for the collective performance in simple tasks for different types of networks. All other model parameters are identical to those considered in Fig.~2 of the main text.

Panel A shows results for random networks that are generated by the configuration model with a constant degree of $k=4$, i.e., with a degree distribution $p(k)=\delta(k-4)$.
In panel B, we consider the case of small-world networks, generated by the Watts-Strogatz model.

As for random networks with a Poisson degree distribution (considered in the main text), we find that in both cases, i.e., for random networks with fixed degree and small-world networks, diversity is detrimental in simple tasks: the time to reach optimal performance increases with diversity.

\subsection*{Complex tasks}
\subsubsection*{Performance on a fully connected network}

Figure~\ref{fig:fullyConnected} shows the collective performance over time obtained on a fully connected network for a complex task. Different colors correspond to different levels of diversity. 

In the main text, we have reported that in complex tasks the benefits of diversity increase for increasing link densities of the underlying random networks. 
Here we show that the result holds in the limit of a fully connected network: final collective performance increases with increasing levels of (skill) diversity.

\subsubsection*{Varying the system size}
\begin{figure}\begin{center}\includegraphics[width=\linewidth]{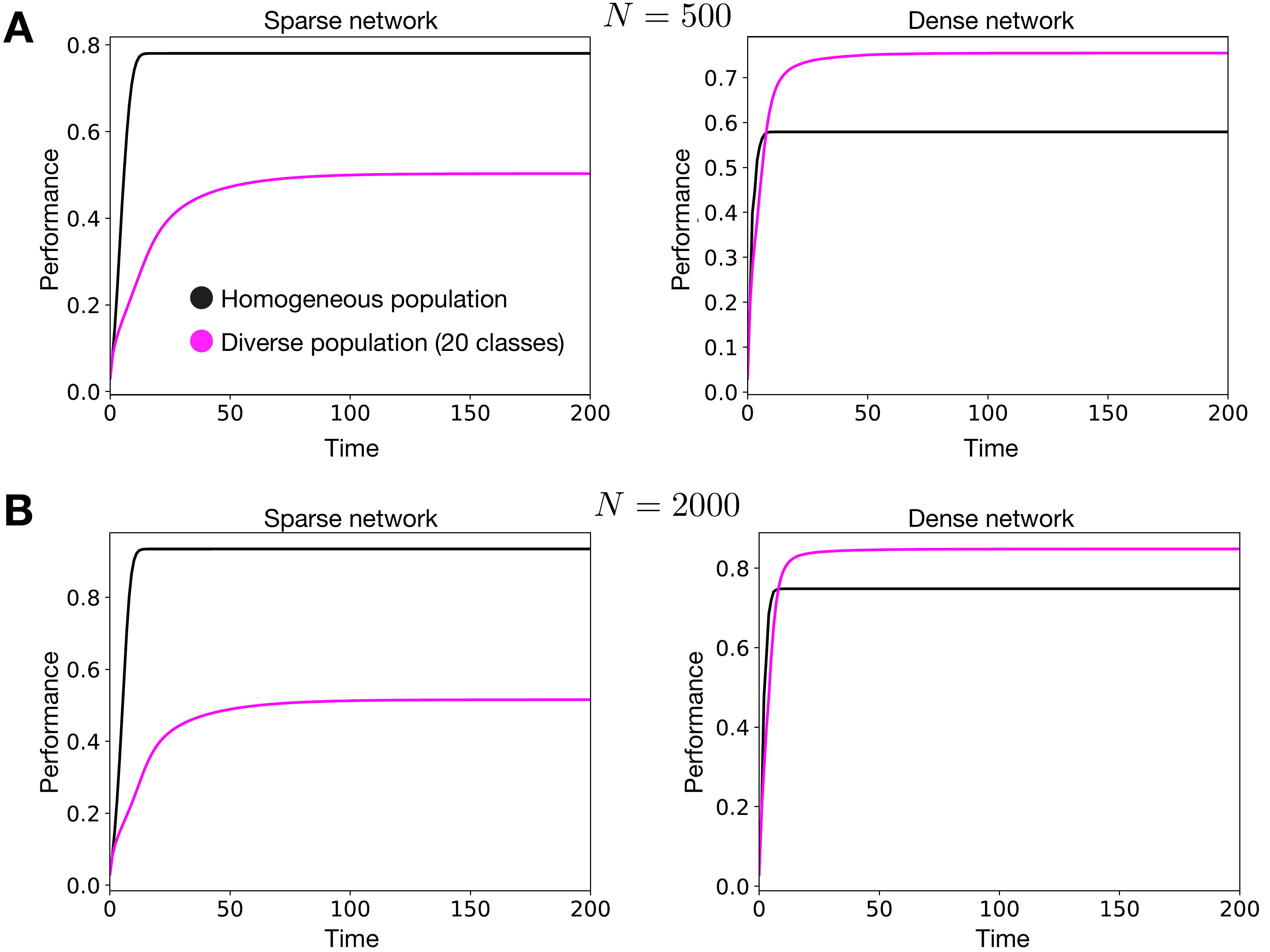}
\caption{Collective performance in complex tasks for $N=500$ (panel A) and $N=2000$ (panel B). The average degrees of the underlying random networks are identical to those of Fig.~3 (panels A and B) of the main text, i.e., we set $\langle k\rangle = 4$ and $\langle k\rangle = 40$ for sparse and dense networks, respectively. The results are averaged over 2500 realizations.}
\label{fig:figureSM_WS_varyN}
\end{center}
\end{figure}

In the main text, we considered populations of size $N=1000$. In Fig.~\ref{fig:figureSM_WS_varyN}, we show additional results for different system sizes.

In particular, we reproduced Fig.~3 (panels A and B) of the main text with half ($N=500$, Fig.~\ref{fig:figureSM_WS_varyN}A) and double the system size ($N=2000$, Fig.~\ref{fig:figureSM_WS_varyN}B). 
As expected, we find that the final values of the collective performance differ slightly. 
In general, higher values of $N$ lead to better performance, due to the fact that large populations can sample the payoff landscape more thoroughly than small populations.

More importantly, however, the qualitative behavior reported in the main text is recovered. 
In both cases, for smaller (panel A) and larger (panel B) system sizes, diverse populations outperform homogeneous ones on dense networks, while the opposite is true for sparse networks.

\subsubsection*{Different network models}
\begin{figure}\begin{center}\includegraphics[width=\linewidth]{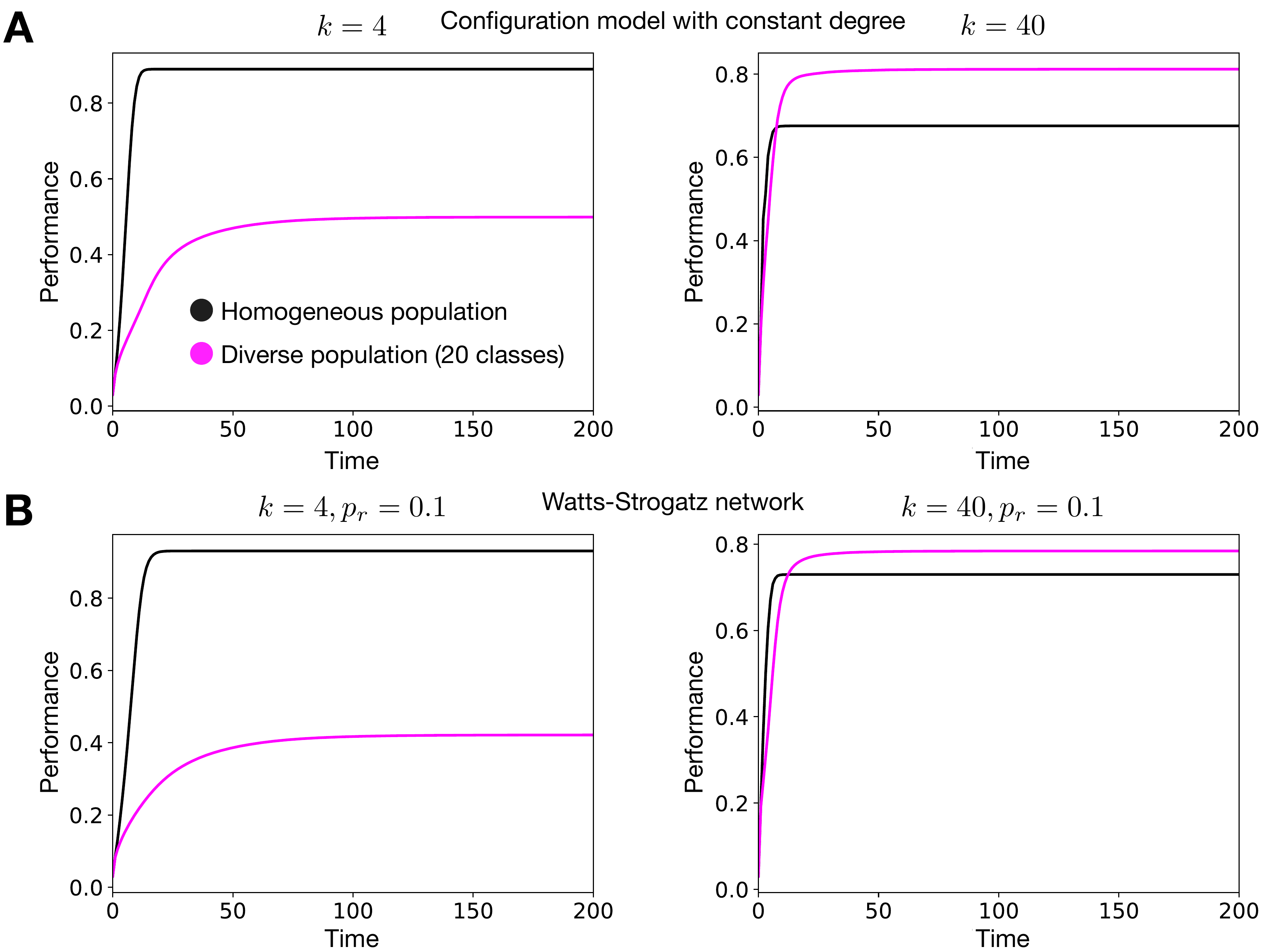}
\caption{Collective performance in complex tasks. Panel A shows results for networks generated by the configuration model with a constant degree for each node: $k=4$ (left) and $k=40$ (right). 
In panel B, we show results for small-world networks generated by the Watts-Strogatz model, where each link of a regular grid network is rewired with probability $p_r$. 
Here we chose a regular gird network where each node is connected to its $k=4$ (left) and $k=40$ (right) nearest neighbors and set the rewiring probability to $p_r=0.1$. In all shown cases the system size is $N=1000$. 
The results are averaged over 2500 realizations.}
\label{fig:figureSM_different_network_models}
\end{center}
\end{figure}

\begin{figure}\begin{center}\includegraphics[width=0.7\linewidth]{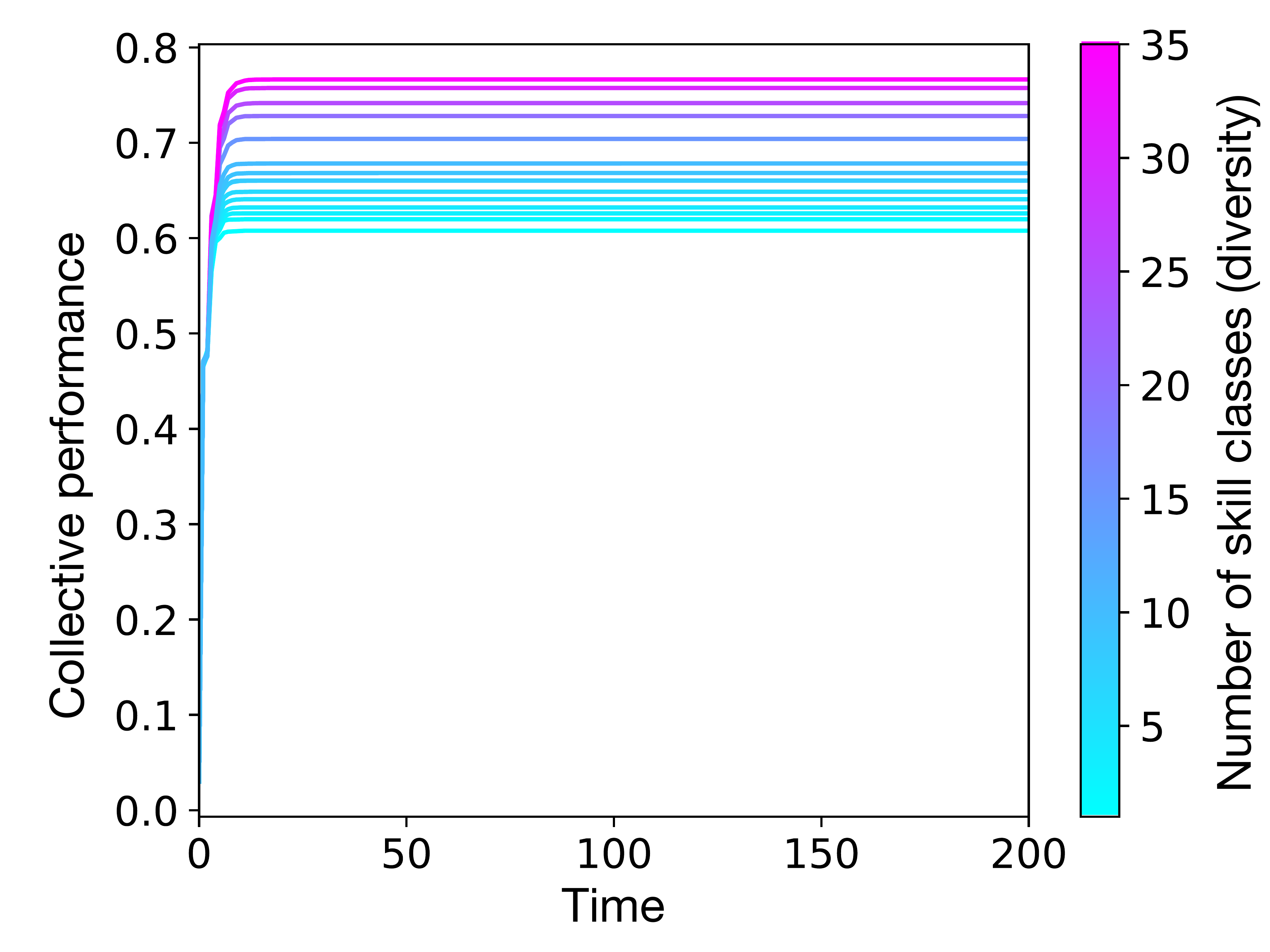}
\caption{Collective performance over time for populations with increasing levels of diversity (number of skill classes) on a fully connected network. The colors of the trajectories encode the number of skill classes in populations of $N=1000$ agents.
The results are averaged over 2500 realizations.}
\label{fig:fullyConnected}\end{center}\end{figure}

In Fig.~\ref{fig:figureSM_different_network_models}, we reproduced Fig.~3 (panels A and B) of the main text for different types of networks, while fixing all other model parameters.

In panel A, we show results for random networks generated by the configuration model with fixed degree $k$, which are either sparse ($k=4$, left) or dense ($k=40$, right).
In panel B, we consider the case of sparse ($\langle k \rangle = 4$) and dense ($\langle k \rangle = 40$) small-world networks.

In both cases our main conclusion holds: link density modifies the effect of diversity, and diverse populations outperform homogeneous ones in densely connected networks.

\end{document}